# Optimizing Beam Transport in Rapidly Compressing Beams on the Neutralized Drift Compression Experiment - II

Anton D. Stepanov (University of Washington)
Erik P. Gilson, Igor D. Kaganovich (Princeton Plasma Physics Lab)
Peter A. Seidl, Arun Persaud, Qing Ji, Thomas Schenkel (Lawrence Berkeley Lab)
Alex Friedman, John J. Barnard, David P. Grote (Lawrence Livermore Lab)

**ASTRACT**

The Neutralized Drift Compression Experiment-II (NDCX-II) is an induction linac that generates intense pulses of 1.2 MeV helium ions for heating matter to extreme conditions. Here, we present recent results on optimizing beam transport. The NDCX-II beamline includes a 1-meter-long drift section downstream of the last transport solenoid, which is filled with charge-neutralizing plasma that enables rapid longitudinal compression of an intense ion beam against space-charge forces. The transport section on NDCX-II consists of 28 solenoids. Finding optimal field settings for a group of solenoids requires knowledge of the envelope parameters of the beam. Imaging the beam on scintillator gives the radius of the beam, but the envelope angle $dr/dz$ is not measured directly. We demonstrate how the parameters of the beam envelope ($r$, $dr/dz$, and emittance) can be reconstructed from a series of images taken at varying B-field strengths of a solenoid upstream of the scintillator. We use this technique to evaluate emittance at several points in the NDCX-II beamline and for optimizing the trajectory of the beam at the entry of the plasma-filled drift section.

**INTRODUCTION**

NDCX-II is a 10-meter-long pulsed induction ion accelerator that produces short (2 - 30 ns FWHM) intense pulses of 1.2 MeV helium ions. Presently, the device is capable of delivering a fluence of 0.7 J/cm$^2$ and is studying radiation damage in materials [Seidl16]. In parallel, an effort to tune the accelerator to increase fluence on target is underway.

The NDCX-II beamline is illustrated in Figure 1. The helium beam is extracted from a multicusp filament-driven plasma ion source [Ji14] at an initial energy of 135 keV. As the ion bunch travels through the beamline, it passes through 12 induction cells that accelerate the beam to a final energy of 1.2 MeV. Besides accelerating the beam, the induction cells are designed to apply a head-to-tail velocity tilt to the ion bunch, i.e. the head of the bunch is decelerated and the tail is accelerated. This results in longitudinal compression of the bunch and a corresponding increase in beam current and line charge density.

The 12 accelerating induction cells are embedded in a 28-solenoid transport lattice. After the transport lattice, the beam passes through a 1 meter long plasma-filled drift column where it undergoes the final longitudinal compression stage before the target. The plasma is generated by a Ferroelectric Plasma Source (FEPS) [Gilson14]. The plasma neutralizes the space charge of the beam [Kaganovich10], which enables a high degree (~x10) of longitudinal compression. Immediately downstream of the FEPS, the beam enters an 8-Tesla 10 cm long Final Focus Solenoid (FFS), which focuses the beam onto the target. The bore radius of the final focus solenoid is small (R = 2 cm) compared to the radius of the beam pipe in the accelerator (4 cm). Passing the beam through the small bore of the FFS with minimal scraping losses is a significant challenge, as we describe later.

The NDCX-II project pushes the capabilities of induction linac technology to develop a compact, low-cost approach to generating extremely high ion beam fluence with short (ns) pulse duration. The beam dynamics on NDCX-II is inherently complex for a number of reasons. The successive applications of the longitudinal velocity tilt result in growing complexity of the longitudinal phase-space of the beam. This velocity spread affects transverse dynamics because the focusing strength of the transport solenoids is a function of the particle velocity. The effect of (generally nonlinear) space charge forces is further complicated by the fact that both the bunch current and energy increase during propagation, resulting in non-monotonic variation of the beam perveance $Q \propto I/V^{3/2}$. Lastly, some sections of the accelerator are filled with plasma to neutralize the space-charge of the beam, which rapidly reduces the self field of the beam and introduces further complexity.

As the result of these factors, source-to-target simulations can easily diverge from experimental reality. This especially concerns transverse beam parameters, such as radius and angle of the beam envelope (*r*, *dr/dz*). Direct measurements of the transverse phase space distribution are difficult due to limited diagnostic access in a crowded lattice of a compact accelerator. However, reliable knowledge of the envelope parameters is often necessary for tuning the solenoid lattice. The previously-mentioned problem of optimizing the trajectory of the beam in the 1-meter long plasma-filled drift section is but one example.

While tackling these issues on NDCX-II, scintillator imaging has emerged as a powerful and flexible diagnostic technique. Reliance on scintillators (instead of 2-slit emittance scanners, for instance) is largely the result of practical concerns. Scintillators can be inserted into the beam with minimal (few cm) longitudinal "real estate" requirements. A single intensified CCD camera positioned at the downstream end of the accelerator can image scintillators at several *z*-locations to measure the transverse current density *j(x,y)* of the beam. Nonetheless, most of the beamline is inaccessible to direct measurements due to the limited number of diagnostic access ports. Furthermore, the envelope angle *dr/dz* cannot be measured directly without inserting additional hardware (such as a movable slit in front of the scintillator plane) into the beam.

The amount of useful information generated by the diagnostics can be increased by measuring the response of the system to its controls. For instance, the spot size of the beam can be measured as a function of solenoid field strength. Then, an inverse problem can be formulated: given some measured dependence of beam radius on solenoid strength *R(B)* at *z = L*, what are the parameters of the beam envelope (*r*, *dr/dz*) at *z = 0*? Solving this problem requires defining a model to calculate the experimentally-measurable quantities as a function of the unknown variables. Then, unknown model parameters can be found by numerical optimization methods.

In the present article, we describe the technique developed on NDCX-II for reconstructing beam parameters inaccessible to direct measurement. The reconstruction technique is based on measuring the spot size of the beam as a function of solenoid strength. Extracting an effective beam radius from the data is accomplished by identifying and exploiting self-similarity in the scintillator images. An envelope model with 3 unknown parameters (beam radius, angle, and perveance) is matched to the data by a particle-swarm optimization algorithm. The validity of reconstructed parameters has been confirmed through agreement with other diagnostics. Our reconstruction technique is similar in spirit to the well-known "solenoid scan" approach to measuring emittance, where emittance is determined from the minimum beam radius downstream of a solenoid lens. However, in contrast with previous work on this subject (e.g. [Poorrezaei13]), the complete shape of the radius vs. B-field curve is taken into account instead of 1 or 2 points. In addition, no measurements of the beam divergence angle are required.

**RECONSTRUCTION TECHNIQUE**

This reconstruction technique presented here was developed to address practical issues encountered on NDCX-II while tuning the accelerator for higher performance. Two such issues, and their resolution via the reconstruction method, are described in this article. These are intended to serve as examples to illustrate the general approach and its potential utility for experimental accelerator physics.

The first issue concerns optimizing the trajectory of the beam through the FEPS to minimize scraping on the 4-cm diameter entrance aperture of the final focusing solenoid (FFS). This requires finding optimal settings for the 3 transport solenoids immediately upstream of the FEPS (#26-#28). Within the assumptions of the envelope model, scraping may be attributed to the effect of large beam emittance, perveance, or beam centroid offset. Otherwise, a setting of the last transport solenoid (#28) could be found that matches the beam into the FFS without scraping for any reasonable beam envelope parameters ($r, dr/dz$) at the entrance of solenoid #28. In the case of emittance (or perveance) limited transport, it is desired to maximize the radius of the beam at the entry of SRK28. This will reduce the minimum attainable beam radius at the FFS entry, and, correspondingly, decrease scraping losses. The described reconstruction technique makes it possible to infer the radius of the beam and its divergence angle at the entrance of solenoid #28 from a measurement of beam radius on scintillator downstream of the solenoid at several solenoid strengths.

The second issue concerns uncertainty about the initial emittance of the beam produced by the NDCX-II ion source and injector, which, again cannot be measured directly due to limited diagnostic access. The same approach is applied to infer the source emittance from a measurement of beam radius on scintillator as a function of transport solenoid strength a few lattice periods downstream of the injector. A general formulation of the inverse problem and the method for solution is given below.

*Inverse problem*

The inverse problem can be described as follows. A beam with an unknown initial radius and divergence angle at $z = 0$ passes through a solenoid lens with a known magnetic field profile. The magnetic field strength of the solenoid can be varied. The radius of the beam versus the magnetic field of the solenoid $R_L(B)$ is measured on a screen located at $z = L$. Given the measured $R_L(B)$, what are the initial radius $r(z = 0)$ and divergence angle $r'(z = 0)$?

In general, solving an inverse problem requires a model relating the unknown parameters, in this case $R_0$ and $R_0'$, to measured quantities. We make the assumption that the experimental beam obeys the RMS envelope equation:

$$\frac{d^2 R}{dz^2} = -k(z)^2 R + \frac{Q}{R} + \frac{\epsilon^2}{R^3}$$

Here $k(z) = eB(z)/\sqrt{8U_b M_i}$ is the focusing strength of the solenoid with magnetic field $B(z)$, $e$ is the electron charge, $M_i$ is the ion mass, and $U_b$ is the kinetic energy of the ions. The parameter $Q = I_B \sqrt{M_i}/4\pi\epsilon_0 \sqrt{2e} V_B^{3/2}$ is the dimensionless perveance and $\epsilon$ is the 4x-RMS unnormalized emittance, and R and R' are the 2xRMS beam radius and divergence angle. The perveance $Q$, which can be determined from the beam current and energy, is assumed to be known and constant from $z = 0$ to $z = B$. The emittance $\epsilon$ is treated as an unknown constant parameter alongside the initial radius and divergence angle $(R_0, R'_0)$. Thus, given $(R_0, R'_0, \epsilon)$, the envelope equation can be solved numerically for the beam radius $R_L$ at $z = L$ for a magnetic field $B$ in the solenoid lens. For $B(z)$, a hard-edge profile is assumed (i.e. $B(z) = B_0$ for $L1 < z < L2$ and zero elsewhere). The effective length of the solenoid $l = L2 - L1 = 16.6$ cm is determined based on experimental measurements of $B(z)$.

Figure 2 shows examples of calculated envelope trajectories $R(z)$ at different values of the solenoid magnetic field. The relative positions of the solenoid and the screen $(z = L)$ correspond to the experimental location of the scintillator downstream of the last transport solenoid. Given an initial set of beam parameters $(R_0, R'_0, \epsilon)$, this model can be used to calculate the radius of the beam on scintillator as a function of solenoid field $B$, denoted as $R_L^{env}(B \mid [R_0, R'_0, \epsilon])$. The inverse problem can be solved by formulating an optimization problem to find $[R_0, R'_0, \epsilon]$ that minimizes the difference between the measured and calculated beam radius as a function of $B$. For this, the following "error function" is used:

$$J(R_0, R_0', \epsilon) = \sum_{i=0}^{N} \left( \frac{R_L^{meas}(B_i) - R_L^{env}(B_i \mid [R_0, R_0', \epsilon])}{R_L^{meas}(B_i)} \right)^2$$

Since $R_L^{meas}(B)$ is measured for $N$ values of the magnetic field $B_i$, a discrete sum is used in the above expression.

With a suitable definition of beam radius vs. $B$ field from scintillator data, a minimum of the error function $J(R_0, R_0', \epsilon)$ can be found by numerical optimization methods. The algorithm used in this work is Particle Swarm Optimization [Kennedy11], although other methods are expected to work as well. A particular advantage of Particle Swarm Optimization is the simplicity of implementation. The algorithm does not rely on calculating gradients of the input function, so the output of any numerical calculation can be used to define $J$, such a numerical solution to the envelope equation modeling the experimental lattice. Extending the optimization scheme to higher dimensions (by letting perveance $Q$ be a free parameter, for example) is straightforward as well.

*Defining beam radius from experimental data*

Evaluating the error function $J(R_0, R_0', \epsilon)$ requires extracting values for the beam radii from scintillator images taken at different solenoid strengths. A measure that is commonly used is $R_B = 2R_{RMS}$, which has the advantage of corresponding to the hard edge of the uniform or KV distribution [Lund09]. However, the $R_{RMS}$ measure of experimental beam profiles often does not converge due to the presence of wide "tails" (note that analytically, RMS radius of a Lorentzian distribution is undefined). Thus, obtaining RMS radius from a scintillator image often requires subtracting some constant background to artificially cut off the distribution, which can make the inferred RMS radius be sensitive to the background assumption. This is especially problematic when a consistent measure of radius is desired for a set of data with significant variation of the intrinsic beam radius. Since integrated fluence $\int_0^\infty j(r)\, r\, dr \simeq \text{const}$ for constant beam current, the peak brightness decreases with increasing beam radius. Subtracting a constant background to obtain RMS radius can result in significant inconsistency in how the radius is defined between profiles with small and large radii.

On NDCX-II, we found that the beam profiles $j(r)$ were reasonably self-similar under a transformation that "stretches" the profile by a scalar magnification factor $M$:

$$j(r) \rightarrow j(r \cdot M)/M^2$$

This transformation preserves the total fluence: $2\pi \int j(r) \cdot r \cdot dr = \text{const}$. Figure 3a plots a set of transformed profiles illustrating their self-similarity. To find the radii for a set of self-similar profiles, it is sufficient to define the radius $R_0$ for a single profile, to which a value $M = 1$ is assigned. For the remaining profiles, $M(B)$ can be easily found from the data by taking the square root of the ratio of the peak intensity with the peak intensity of the $M = 1$ profile. Radius can then be found as $R(B) = R_0 M(B)$, as illustrated in Figure 3b.

## APPLICATION TO NDCX-II DATA

*Case 1: finding optimal settings for the last 3 transport solenoids*

At the end of the NDCX-II transport lattice, the beam is launched through a 1-meter-long plasma-filled drift section. In the drift section, no transverse focusing forces are applied. Since the space-charge of the beam should be well-neutralized by the plasma, the beam is expected to propagate ballistically, with its trajectory set by the envelope parameters at the exit of the transport lattice. The envelope parameters (radius and divergence angle) can be controlled by tuning the magnetic field strengths of the final group of solenoids of the lattice. At the end of the 1-meter long drift section, the beam enters the Final Focus Solenoid with a small (2 cm) bore radius. In the experiment, it was found that significant particle loss of the beam at the entry or upstream of the Final Focus Solenoid occurred, leading to losses of charge on target. Thus, it was necessary to find optimal $B$-field values for the last three solenoids (#26-28) that minimize scraping losses.

In the framework of the envelope model, particle loss corresponds to the beam radius at the entrance of the Final Focus Solenoid being greater than the 2 cm bore radius. Since the space-charge of the beam was expected to be well-neutralized, scraping was attributed to finite beam emittance. Chromatic aberration due to the intrinsic beam velocity spread on NDCX-II was also considered, but the effect was estimated to be too small to explain the measured scraping losses. For an emittance-dominated and monochromatic beam, the minimum attainable radius on target decreases with increasing initial radius. This is evident from the envelope equation, which can be solved exactly for the case of $Q = 0$:

$$R(z) = \sqrt{R_0^2 + 2R_0 R_0' z + \left(\epsilon^2/R_0^2 + R_0'^2\right)z^2}.$$

The above equation gives the radius of the beam on target at $z = L$ as a function of the initial envelope parameters at $z = 0$. One can find the minimum of $R(z = L)$ with respect to the initial angle $R_0'$ by solving $dR(L)/dR_0' = 0$. This yields a minimum radius $R(L)_{min} = \epsilon L/R_0$ with $R'_0 = -R_0/L$.

Since the minimum attainable radius is inversely proportional to the initial radius $R_0$, it was desired to tune the last three transport solenoids so the beam enters the drift section with a radius as close to the 4 cm transport radius as possible, and with a divergence angle $R'_0 = R_0/L$.

In lieu of a direct divergence angle measurement, the reconstruction technique described previously was applied to infer the envelope parameters from measurements of beam radius versus the magnetic field in the last transport solenoid (#28). The scintillator was positioned 34 cm downstream of the exit of solenoid #28. The experimental arrangement of this measurement is shown in Figure 4. The envelope parameters are reconstructed at the entry of the solenoid, which is taken to correspond to the point $z = 0$. Given the beam envelope parameters at $z = 0$, the envelope model can be used to solve for the radius and divergence angle of the beam at the exit of solenoid #28, making it possible to determine whether the beam is on an optimal trajectory through the drift section. If not, adjustments are made to the upstream solenoids (#27 and #26), and the scan of solenoid #28 is repeated.

First, it was necessary to find the unknown emittance of the beam. In order to determine a unique value of $\epsilon$, it was found that the measured $R_L^{meas}(B)$ curve has to pass through a minimum, i.e. the radius of the beam has to begin increasing with solenoid field. By increasing the strengths of solenoids #26 and #27 above their standard settings, a measurement of $R_L(B)$ shown in Figure 5a was produced, which passes through a minimum. By applying the envelope reconstruction algorithm to this data, the emittance of the beam was determined to be $\epsilon = 5.2\text{e-}2$ cm-rad. This value was found by solving the 3-D optimization problem, where the initial radius, divergence angle, and emittance, are unknown. To confirm that the found value of emittance is indeed unique, Figure 5b plots the minimum attainable error vs. choice of $\epsilon$ for the 2D optimization problem. One can see the presence of a minimum in the error at $\epsilon = 5.2\text{e-}2$ cm-rad.

Once the emittance of the beam has been determined, the reconstruction algorithm was applied to another set of measurements of $R_L(B)$ with decreased (standard) B-field values in solenoids #26 and #27. This data is shown in Figure 3b. Using the reconstruction algorithm, we can determine the trajectory of the beam at the exit of solenoid #28, i.e. at the entrance of the drift section. The radius of the beam was 3 cm and the convergence angle was 0.043 rad. According to the analytic formula derived previously, the optimal convergence angle would be $R'_{opt} = R_0/L \simeq 3\text{cm}/100\text{cm} \sim 0.03$ rad. However, testing the effect of further tuning with the envelope model showed that these experimental settings were close to optimal, with the main cause of scraping being due to large beam emittance.

The unnormalized measured emittance $\epsilon = 5.2\text{e-}2$ cm-rad corresponds to a normalized emittance $\epsilon_{norm} = \beta\epsilon = 12$ mm-mrad. Since the beam undergoes multiple acceleration "kicks" on NDCX-II, the normalized emittance serves as a useful metric for comparing the beam emittances at different locations in the beamline. Since the large measured emittance resulted in significant loss of charge on target, it became necessary to determine why the emittance of the beam is so high and whether or not it can be reduced.

*Case 2: measuring the plasma ion source emittance*

Given the large value of emittance measured at the end of the transport lattice discussed in the previous section, it motivated us to attempt to determine the origin of the high beam emittance. In order to infer the initial emittance of the beam at the exit of the ion source, a scintillator was installed 134 cm downstream of the source, after the third solenoid in the accelerator. The envelope reconstruction technique was applied to measure the source emittance in a similar manner to the previous section.

After exiting the source, the ion beam passed through the first three NDCX-II transport solenoids before it reached the scintillator screen. The experimental setup is shown schematically in Figure 6a. The B-field in the first transport solenoid (#1) was set at 0.8 Tesla. A lower magnetic field strength resulted in significant losses of beam current at the measurement location. The field in the second transport solenoid (#2) was varied for the scan. The B-field in solenoid #3, which was located immediately upstream of the scintillator, was set to zero. This provided for the drift distance so the effects of solenoid #2 on the beam trajectory would be manifested.

The measured beam radius versus the B-field in solenoid #2 is shown in Figure 6b, together with the result of the envelope reconstruction routine. Note that for this data set, the shapes of the measured beam profiles at B = 0 Tesla was not self-similar with the profile shapes at nonzero B-field values. Thus, these data points were excluded from the inputs to the reconstruction algorithm. For this measurement, the beam energy was 150 keV and the beam current was 40 mA, corresponding to a perveance $Q$ ~ 1e-3. The inferred unnormalized emittance of the beam was $\epsilon = 1.4$e-1 cm-rad. The corresponding normalized emittance was $\epsilon_{norm} = 12$ mm-mrad, which is identical to the value of $\epsilon_{norm}$ measured at the entry of the drift section. This suggests that the origin of the high beam emittance on NDCX-II may be due to a higher-than-expected emittance of the beam from the ion source and injector.

## **CONCLUSIONS**

We formulate an inverse problem approach to reconstructing beam phase space in an accelerator experiment. It is shown that 3 beam envelope parameters – radius, divergence angle, and emittance – can be deduced from a measurement of beam radius vs. solenoid strength with a numerical optimization algorithm. This simple and general technique can be applied on other experiments operating with beams that can be reasonably well-described by the envelope equation. The main benefit of this approach is that beam parameters that cannot be measured directly can be inferred with sufficient confidence.

The numerical reconstruction technique was developed on the NDCX-II accelerator, which has intrinsically complex beam dynamics due to simultaneous beam compression and acceleration. This complexity makes NDCX-II a good platform for investigating the utility of the inverse problem approach. Based on our investigation, the main factor limiting the fluence on target on NDCX-II is the operation of the ion source and injector, which may be producing a beam pulse with much higher normalized emittance (12 mm-mrad normalized) than was expected (~2 mm-mrad normalized). High emittance limits the radial compressibility of the beam and results in charge losses due to scraping of the beam on the walls of the accelerator.

Reducing the emittance of the ion source can dramatically improve the performance of the NDCX-II accelerator. Several approaches can be attempted towards that end. The ion source itself has a number of "knobs," including the voltages on the extraction and suppressor grids, as well the voltage ratios in the 135 kV injector. These control parameters of the ion source can be optimized in an effort to reduce the source emittance. With improved ion source performance, a significant increase in target fluence can be readily expected on NDCX-II, potentially enabling targets to be heated to warm dense matter conditions.

**AKNOWLEDGEMENTS**

This work is supported by the Office of Science of the US DOE under contracts DE-AC0205CH11231, DE-AC52-07NA27344, and DE-AC02-09CH11466.

# REFERENCES

[Gilson14] "Ferroelectric Plasma Sources for NDCX-II and Heavy Ion Drivers," E. P. Gilson, *et al*. Nuclear Instruments and Methods in Physics Research A **733**, 226 (2014).

[Ji16] "Development and testing of a pulsed helium ion source for probing materials and warm dense matter studies," Q. Ji, P. A. Seidl, W. L. Waldron, J. H. Takakuwa, A. Friedman, D. P. Grote, A. Persaud, J. J. Barnard, and T. Schenkel. Review of Scientific Instruments **87**, 02B707 (2016).

[Kaganovich10] "Physics of Neutralization of Intense High-Energy Ion Beam Pulses by Electrons," I. D. Kaganovich, R. C. Davidson, M. A. Dorf, E. A. Startsev, A. B Sefkow, A. F. Friedman and E. P. Lee, Physics of Plasmas **17**, 056703 (2010).

[Kennedy11] "Particle swarm optimization," J. Kennedy, Encyclopedia of machine learning, Springer, 760 (2011).

[Lund09] "Generation of initial kinetic distributions for simulation of long-pulse charged particle beams with high space-charge intensity," Physical Review Special Topics Accelerators and Beams **12**, 114801 (2009).

[Seidl16] "Recent Experiments at NDCX-II: Irradiation of Materials Using Short, Intense Ion Beams," P. A. Seidl, Q. Ji, A. Persaud, E. Feinberg, B. Ludewigt, M. Silverman, A. Sulyman, W. L. Waldron, T. Schenkel, J. J. Barnard, A. Friedman, D. P. Grote, E. P. Gilson, I. D. Kaganovich, A. Stepanov, F. Treffert, M. Zimmer.

[Seidl17] "Irradiation of materials with short, intense ion pulses at NDCX-II," P. A. Seidl, J. J. Barnard, E. Feinberg, A. Friedman, E. P. Gilson, D. P. Grote, Q. Ji, I. D. Kaganovich, B. Ludewigt, A. Persaud, C. Sierra, M. Silverman, A. D. Stepanov, A. Sulyman, F. Treffert, W.L. Waldron, M. Zimmer, and T. Schenkel, Laser and Particle Beams **35**, 373 (2017).

[Poorrezaei13] "New technique to measure emittance for beams with space charge," K. Poorrezaei, R. B. Fiorito, R. A. Kishek, and B. L. Beaudoin. Physical Review Special Topics - Accelerators and Beams **16**, 082801 (2013).

**FIGURES**

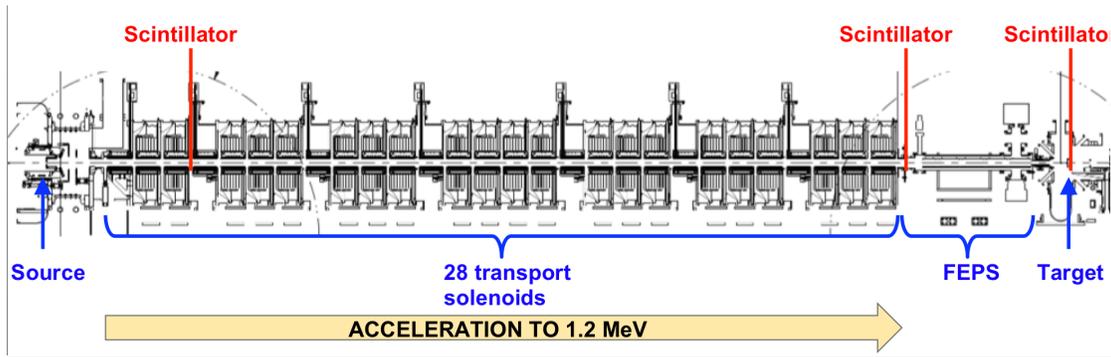

Figure 1: NDCX-II beamline. The accelerator is 10 meters long from source to target. The He$^+$ ion beam is extracted from a multicusp plasma ion source and transported through a 28-solenoid lattice towards the Ferroelectric Plasma Source (FEPS). Inside the FEPS, a volume plasma is generated that neutralizes the space-charge of the beam and enables longitudinal and transverse compression of the ion pulse. The locations where scintillator measurements of beam spot size were taken are indicated in the Figure.

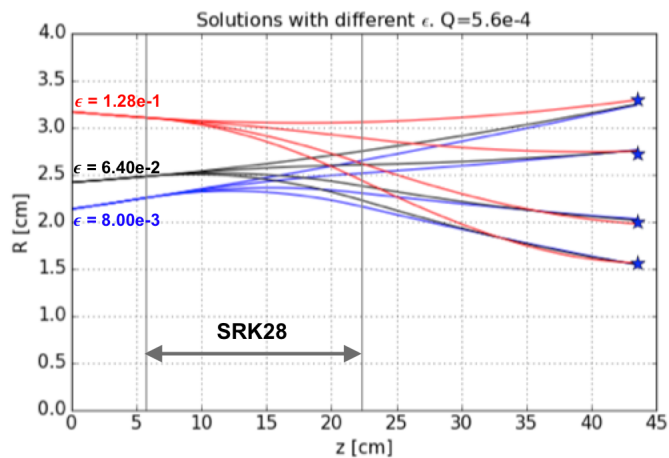

Figure 2: Example plots of a set of envelope trajectories through the last transport solenoid upstream of the FEPS.

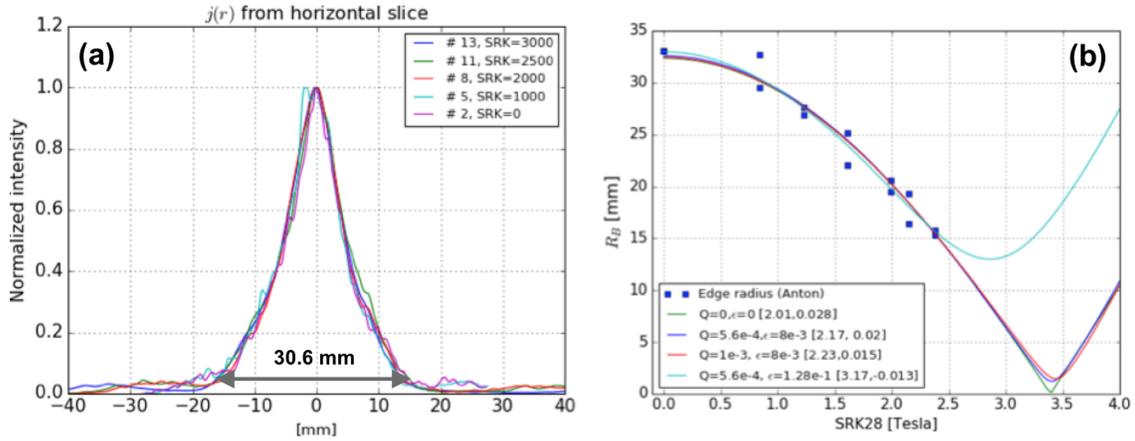

Figure 3: (a) Beam profiles *j(r)* measured on scintillator are self-similar under the transformation $j(r) \to j(r \cdot M)/M^2$ where $M$ is a scalar magnification factor. (b) Measured radius vs. $B$ field. The radius is defined based on the factor $M$ of the self-similar transformation as $R(B) = M(B) \cdot R_0$.

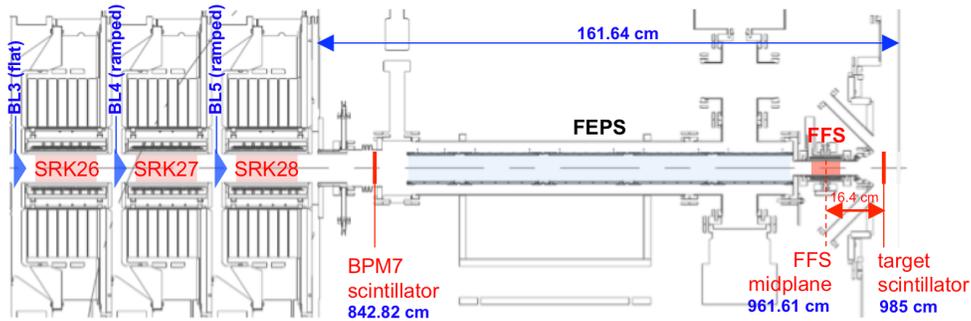

Figure 4: Geometry of the measurement to infer the beam envelope parameters at the entry top the last transport solenoid immediately upstream of the FEPS.

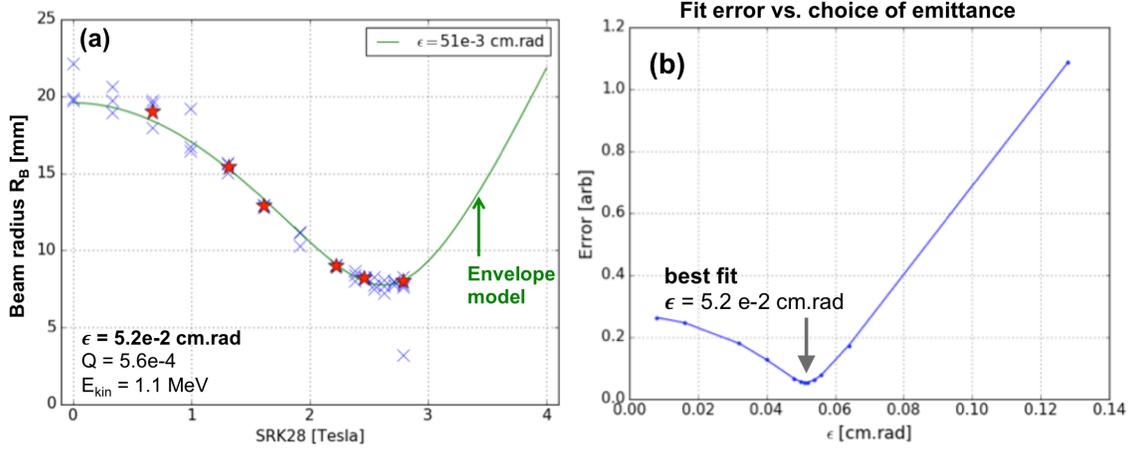

Figure 5: (a) Measured $R_L(B)$ at increased fields in solenoids #26 and #27. (b) Fit error versus choice of emittance, showing a minimum at $\epsilon = 5.2$e-2 cm-rad.

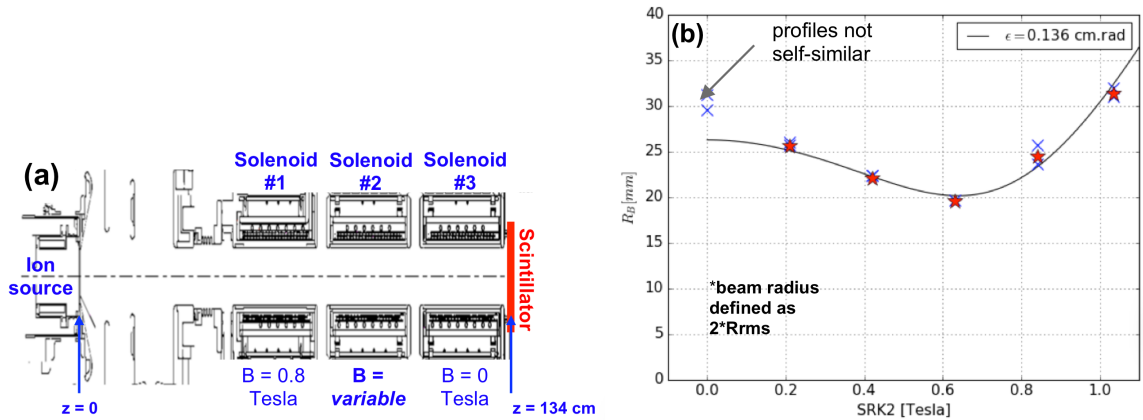

Figure 6: (a) Setup of the source emittance measurement. The scintillator was placed 134 cm downstream of the ion source. Solenoid #1 was set at 0.8 Tesla to direct ion current from the source into the accelerator. The field in solenoid #2 was varied for the scan, while solenoid #3 was turned off to give the beam some drift distance; (b) Measured beam radius vs. the B-field in solenoid #2.